\documentclass[12pt]{article}
\usepackage{amsmath}
\usepackage{graphicx}
\usepackage{amsfonts}
\usepackage{amssymb}

\begin{document}

\bigskip\pagenumbering{arabic} \thispagestyle{empty} \setcounter{page}{0}
\begin{flushright}
ZU--TH 41/99
\end{flushright}

\vfill

\begin{center}

{\Large Reordering of Baryon Chiral Perturbation Theory*\\[60pt]}

{Martin Moj\v{z}i\v{s}$^{a}$ and Joachim Kambor}$^{b}$ \\[30pt]

$^{a)}$ Department of Theoretical Physics, Comenius University\newline
Mlynsk\'{a} dolina, SK--84248 Bratislava, Slovakia \\[8pt]

$^{b)}$ Institut f\"{u}r Theoretische Physik, Universit\"{a}t Z\"{u}%
rich\newline Winterthurerstr. 190, CH--8057 Z\"{u}rich, Switzerland \\[40pt]
\end{center}

\vfill

\begin{abstract}
Reordering of the chiral perturbation series, proposed recently by Becher and
Leutwyler in the framework of $SU(2)$ baryonic ChPT, is applied to the $SU(3)$
case. This results in improved convergence of the chiral expansion of static
properties of the lowest lying baryon octet, which in most cases is quite
impressive. Finite renormalization of coupling constants and the role it plays
in the interpretation of effective field theories is discussed. Some future
tests of the viability of the scheme are proposed too.
\end{abstract}

\vfill
\noindent{\small * Work supported in part by VEGA grant No.1/4301/97, by
Schweizerischer Nationalfonds and by the EEC-TMR Program, Contract No.
CT98-0169.}\newpage

1. The phenomenology of baryons at low energies is reasonably well described
by the chiral symmetry $SU(3)_{L}\times$ $SU(3)_{R},$ broken spontaneously to
$SU(3)_{V},$ as well as it's explicit breaking by non-vanishing quark masses.
The most celebrated successes of chiral $SU(3)$ symmetry go back to the
sixties. These are now understood as tree level results of chiral perturbation
theory (ChPT) --- the low-energy effective theory of QCD. The main virtue of
ChPT is that it enables calculation of loop corrections to these tree results
in a systematic way. However, the loop-corrections turn out to be typically
too large, thereby corrupting the overall good agreement between theory and
experiment at tree level. The agreement can be restored by adjusting large
counterterms at higher orders, but it is commonly believed that this happens
only at the expense of a slow convergence of the perturbation series.

To illustrate the problem, let us quote the results for mass \cite{BM96} and
magnetic moment of the $\Lambda$-hyperon \cite{MS97}
\begin{equation}%
\begin{array}
[c]{l}%
m_{\Lambda}=767\,(1+0.69-0.77+0.54)\,\mathrm{MeV}\\
\mu_{\Lambda}=-1.31\,(0+1-0.82+0.29)\,\mu_{N}%
\end{array}
\label{example}%
\end{equation}
where $\mu_{N}=\frac{e\hbar}{2m_{p}}$ is the nucleon magneton. Here, as well
as in formulae to follow, the numbers in brackets correspond to contributions
from different chiral orders (from the 1st up to the 4th). The pattern is
characteristic for other quantities, like e.g. weak decay constants, as well
as for all members of the baryon octet. The problem of slow convergence of the
static baryon properties was recently studied in \cite{DHB98}, where a remedy
called long distance regularization (a variant of cut-off regularization) was proposed.

It is, however, questionable whether the direct comparison of various
contributions in the chiral expansion of a given observable is the best test
of the rate of convergence. ChPT is, after all, the tool for solving Ward
identities imposed by chiral symmetry, i.e. it \textit{relates} physical
observables which would be independent in the absence of the symmetry. It is
therefore more natural to form an opinion about the rate of convergence from
changes in such relations, at each order of the chiral expansion, rather than
from expansions like (\ref{example}). This point of view was recently emphasized
in \cite{McGBi99} and \cite{BL99}. In \cite{McGBi99} the order $p^5$ correction 
to the nucleon mass was calculated and it was pointed out that, except for a 
small recoil correction, all of the terms found at that order can be absorbed by re-expressing the order $p^3$ contribution in terms of physical coupling constants,
thereby improving the convergence of the chiral expansion of the nucleon mass.
In \cite{BL99}, the reordering was formulated in a form which allows to judge the
rate of convergence from expansion of individual quantities like (1), but again is
in spirit much closer to the comparison of relations between different physical
observables.

The essence of the reordering is very simple and well known: results are
expressed not in terms of bare coupling constants, but rather in terms of the
physical couplings. This amounts to replacing the bare constants by the
physical ones and shifting the difference to higher orders. For rapidly
converging series, this is just useful cosmetics. For slowly convergent
series, the reordering may be quite significant. As an additional example 
to the above mentioned nucleon mass \cite{McGBi99} let us quote 
the expansion of the $\pi$N $\sigma$-term as given in \cite{BL99}. The shift
due to the loop corrections amounts numerically to $-31$ MeV in the expansion
analogous to (\ref{example}), but only to $2$ MeV in the reordered series.

Why is the reordered expansion preferable? Consider the process of pinning
down values of low energy constants (LECs) from experimental data of
quantities forming a set $A$ and then using these values in results for
quantities forming a set $B$. This is nothing else than relating the
quantities from the sets $A$ and $B$. In other words, the relations between
various quantities are embodied in the values of the LECs. Tree level
relations are characterized by values pinned down at the tree level, etc.
Comparison of relations between various quantities at 1-loop and tree level
corresponds to comparison of values of LECs pinned down at the 1-loop and tree
level. This is what is done in the reordered perturbation series. On the other
hand, expansions like (\ref{example}) are comparing values of different
orders, as given by the fit using the complete result up to the highest order
calculated. In such an expansion, the values of counterterms at lower orders
do not characterize relations between physical observables.

In \cite{BL99} the reordered perturbation series was discussed in the
framework of $SU(2)$ baryonic ChPT. However, it is clear that reordering could
be even more relevant in the $SU(3)$ case. The problem with $SU(3)$ baryonic
ChPT, as stressed in \cite{DHB98}, is that the simplicity evident in the
baryon physics at tree level is lost at the one-loop level. The reordered
perturbation series has an immediate benefit: tree level results are not
changed after loop corrections are included. The aim of this note is to
investigate to what extent can the reordered perturbation series help to cure
the problem of slow convergence in $SU(3)$ baryonic ChPT.\bigskip\ 

2. Reordering of the perturbation series discussed in \cite{BL99} was
reordering between different orders in the \textit{loop} expansion. Expansion
(\ref{example}) and other examples collected in \cite{DHB98} are more detailed
--- they treat separately every chiral order. To be able to investigate the
influence of the reordering of perturbation series on these more detailed
expansions, we slightly generalize the approach of \cite{BL99} and perform
reordering between different orders of the \textit{chiral} expansion.

Formally this is achieved by writing any 1st order LEC $c$ in the form%
\begin{equation}
c=c^{(0)}+\delta c^{(1)}+\delta c^{(2)}+\delta c^{(3)}+\ldots\label{LEC1}%
\end{equation}
any 2nd order LEC $d$ in the form%
\begin{equation}
d=d^{(0)}+\delta d^{(1)}+\delta d^{(2)}+\ldots\label{LEC2}%
\end{equation}
etc. Here the superscript $(n)$ refers to the chiral order (rather than to the
number of loops, as in \cite{BL99}). The standard chiral expansion of the
meson-baryon Lagrangian
\begin{equation}
\mathcal{L}_{MB}=\mathcal{L}_{MB}^{(1)}+\mathcal{L}_{MB}^{(2)}+\ldots
\label{L standard}%
\end{equation}
is then rewritten in the form%

\begin{equation}
\mathcal{L}_{MB}=\underset{\mathrm{1st\;order}}{\underbrace{\mathcal{L}%
_{MB}^{(1)}\mid_{c\rightarrow c^{(0)}}}}+\underset{\mathrm{2nd\;order}%
}{\underbrace{\mathcal{L}_{MB}^{(2)}\mid_{d\rightarrow d^{(0)}}+\mathcal{L}%
_{MB}^{(1)}\mid_{c\rightarrow\delta c^{(1)}}}}+\ldots\label{L reordered}%
\end{equation}

It is now straightforward to reorder the chiral expansion of any quantity. Our
starting point is the Lagrangian
\begin{equation}%
\begin{array}
[c]{lll}%
L_{MB}^{(1)} & = & \operatorname*{Tr}\overline{B}\gamma^{\mu}\left[
i\nabla_{\mu},B\right]  -m\operatorname*{Tr}\overline{B}B\\
& + & \frac{1}{2}d_{A}\operatorname*{Tr}\overline{B}\gamma^{\mu}\gamma
^{5}\left\{  u_{\mu},B\right\}  +\frac{1}{2}f_{A}\operatorname*{Tr}%
\overline{B}\gamma^{\mu}\gamma^{5}\left[  u_{\mu},B\right]
\end{array}
\label{L1}%
\end{equation}%
\begin{equation}%
\begin{array}
[c]{lll}%
L_{MB}^{(2)} & = & b_{0}\operatorname*{Tr}\chi_{+}\operatorname*{Tr}%
\overline{B}B+d_{m}\operatorname*{Tr}\overline{B}\left\{  \chi_{+},B\right\}
+f_{m}\operatorname*{Tr}\overline{B}\left[  \chi_{+},B\right] \\
& + & d_{V}\operatorname*{Tr}\overline{B}\sigma_{\mu\nu}\left\{  F_{+}^{\mu
\nu},B\right\}  +f_{V}\operatorname*{Tr}\overline{B}\sigma_{\mu\nu}\left[
F_{+}^{\mu\nu},B\right]  +\ldots
\end{array}
\label{L2}%
\end{equation}
where $\chi_{+}=4B_{0}\operatorname*{diag}\left(  m_{u},m_{d},m_{s}\right)
+\ldots=2\operatorname*{diag}\left(  M_{\pi}^{2},M_{\pi}^{2},2M_{K}^{2}%
-M_{\pi}^{2}\right)  +\ldots$ and ellipses always stand for terms not needed
in what follows. We refrain from giving the explicit form of the relevant
terms in $L_{MB}^{(3)}$ and $L_{MB}^{(4)}$, they can be found in the quoted
papers \cite{BM96} \cite{MS97}.

To illustrate how the reordering works, let us consider the mass of the
$\Lambda$-hyperon. At tree level the result is straightforward
\begin{equation}
m_{\Lambda}^{\mathrm{tree}}=m-2\left(  2M_{K}^{2}+M_{\pi}^{2}\right)
b_{0}-\frac{4}{3}\left(  4M_{K}^{2}-M_{\pi}^{2}\right)  d_{m}.
\end{equation}
We have replaced the bare meson masses by the physical ones, which does not 
affect the result up to the 4th order. Moreover, this procedure is in spirit 
of the considered reordering. At one-loop level one has
\begin{equation}
m_{\Lambda}=m_{\Lambda}^{\mathrm{tree}}+L_{3}(m,d_{A},f_{A})+L_{4}%
(\ldots),\label{mL}%
\end{equation}
where the explicit form of the loop corrections $L_{3}$\ and $L_{4}$\ can be
found in \cite{BM96}. We just recall that $L_{3}$ is a finite loop correction
(no counterterms of the 3rd order), while $L_{4}$ contains loop as well as
counterterm contributions containing 4th order LECs. The latter as well as
some lower order LECs are represented by the ellipses in (\ref{mL}). Then the
reordered series assumes the form
\begin{equation}%
\begin{array}
[c]{lll}%
m_{\Lambda} & = & m^{(0)}\\
& + & \delta m^{(1)}-2\left(  2M_{K}^{2}+M_{\pi}^{2}\right)  b_{0}^{(0)}%
-\frac{4}{3}\left(  4M_{K}^{2}-M_{\pi}^{2}\right)  d_{m}^{(0)}\\
& + & \delta m^{(2)}-2\left(  2M_{K}^{2}+M_{\pi}^{2}\right)  \delta
b_{0}^{(1)}-\frac{4}{3}\left(  4M_{K}^{2}-M_{\pi}^{2}\right)  \delta
d_{m}^{(1)}+L_{3}(m^{(0)},d_{A}^{(0)},f_{A}^{(0)})\\
& + & \ldots
\end{array}
\end{equation}
where the n-th line represents the n-th order of the reordered series.

Analogous expansions hold for other members of the lowest lying baryon octet.
The numerical value of $m^{(0)}$ is obtained by fitting the experimental
values of baryon masses to the 1st order results. This yields $m^{(0)}=1150$
MeV. Numerical values of $\delta m^{(1)},$ $b_{0}^{(0)}$ and $d_{m}^{(0)}$ are
then obtained by the fit with second order results, using for $m^{(0)}$ the
already fixed value $1150$ MeV. In a similar way, the values of $\delta
m^{(2)},$ $\delta b_{0}^{(1)}$ and $\delta d_{m}^{(1)}$ are determined by the
3rd order fit. Here one uses the values for the LECs $d_{A}$ and $f_{A}$
determined via semileptonic weak decays of baryons. The data are well
reproduced already at the tree level, with $d_{A}^{(0)}=0.75$ and $f_{A}%
^{(0)}=0.5$. At the 4th order a number of LECs enter the result. In
\cite{BM96} these were estimated by the resonance saturation principle. Using
these estimates for the 4th order results one obtains the following reordered
chiral expansion of baryon masses
\begin{equation}%
\begin{array}
[c]{l}%
m_{N}=(1150-209+0+3)\;\mathrm{MeV}\\
m_{\Lambda}=(1150-39+1+8)\;\mathrm{MeV}\\
m_{\Sigma}=(1150+39+0+0)\;\mathrm{MeV}\\
m_{\Xi}=(1150+170+0+0)\;\mathrm{MeV}.
\end{array}
\label{masses}%
\end{equation}
Comparison with (\ref{example}) as well as with expansions of $m_{N}$,
$m_{\Sigma}$ and $m_{\Xi}$ as given in \cite{BM96} shows a tremendous
improvement in the rate of convergence.

At first sight this improvement may look too good. For a long time $SU(3)$
ChPT has had a reputation of being a rather slowly convergent theory, so the
result (\ref{masses}), and especially the zeros in the third and fourth orders
may come as a surprise. However, there is nothing mysterious here. All we have
done is to implement a fitting procedure such that higher orders cannot spoil
a prospective agreement achieved at lower orders. And since there is very good
agreement already at the 2nd order, higher corrections must be tiny. Of
course, the large corrections have not disappeared completely but are now
hidden, e.g. in the expansion of $m$. Note, however, that $m$ is not a
physical observable.

A natural objection is that we have made convergence more rapid by hand, i.e.
by reshuffling a (large) part of the originally second order contribution to
higher orders, where it has cancelled the large loop contributions. The
objection is valid, but this is possible only for a small number of
quantities, equal to the number of LECs to the order considered. In our case
we have four LECs, but two of them, namely $m$ and $b_{0}$ enter only in a
particular combination, so the number is effectively reduced to three. And
since the convergence was improved for four masses, this is already a
nontrivial result. Moreover, one can now study other quantities depending on
the same set of LECs. The question then is, if the convergence of these
quantities in the reordered version of the perturbation series is improved too.\bigskip

3. In case of the baryon magnetic moments the reordering is a consequence of
the expansion (\ref{LEC2}) of LECs $d_{V}$ and $f_{V}$. Performing the fit at
various orders of the chiral expansion (relevant formulae are found in
\cite{MS97} ) one obtains the following reordered chiral expansion of baryon
magnetic moments
\begin{equation}%
\begin{array}
[c]{l}%
\!\mu_{p}=(2.56+0.39-0.15)\;\mu_{N}\\
\!\mu_{n}=(-1.60-0.91+0.60)\;\mu_{N}\\
\!\mu_{\Lambda}=(-0.80+0.36-0.18)\;\mu_{N}\\
\!\mu_{\Sigma^{+}}=(2.56-0.33+0.22)\;\mu_{N}\\
\!\mu_{\Sigma^{-}}=(-0.97-0.39+0.19)\;\mu_{N}\\
\!\mu_{\Sigma^{0}}=(0.80-0.36+0.22)\;\mu_{N}\\
\!\mu_{\Xi^{0}}=(-1.60+0.79-0.45)\;\mu_{N}\\
\!\mu_{\Xi^{-}}=(-0.97+0.45-0.15)\;\mu_{N}.
\end{array}
\label{mu}%
\end{equation}
Here the first number in brackets represents the sum of the first two chiral
orders, and is followed by the third and the fourth order contribution.
Comparison with (\ref{example}) as well as with expansions of other magnetic
moments as given in Eq. (29) of \cite{MS97} shows again a (moderate)
improvement in the rate of convergence. From the point of view of the loop
expansion, the convergence is even better, since the loop contribution
(including counterterms from $L_{MB}^{(3)}$ and $L_{MB}^{(4)}$) is given by
the sum of the last two numbers in brackets, which are always of comparable
size and opposite sign. The loop corrections are typically at the level of
20\% of the tree result (4\% in the best, 31\% in the worst case), which is
quite acceptable.

Let us stress that Eq. (\ref{mu}) is not new and practically equivalent to
Table 1. in \cite{MS97} (as it should be, since this Table 1. just presents
magnetic moments calculated in various orders). Our point is that to judge the
rate of convergence for baryon magnetic moments one should consider results of
\cite{MS97} in the form of their Table 1. (as we have done in this article),
rather than in the form of their Eq. (29) (as was done in \cite{DHB98}). \bigskip

4. The examples discussed so far are the only ones where the complete 1-loop
calculation is available. Our reordering between different chiral orders
enables us to study also those quantities for which results are presently
known only up to the 3rd order. As an illustration we take another example
from \cite{DHB98}, namely baryon axial couplings. Here the 3rd chiral order
1-loop contributions are once again too large \cite{BSW85,B98}. As an
illustration let us quote the expansion of the $\Lambda$ decay constant
\begin{equation}
g_{1}\left[  p\Lambda\right]  \left(  0\right)  =-0.5-0.48\label{axial}%
\end{equation}
A similar pattern is observed in all the other cases considered in \cite{B98}.
The first term in (\ref{axial}) corresponds to the first two chiral orders.
The second term is the 3rd order contribution, however it is not complete. In
this case 3rd order LECs of numerous counterterms contributing to axial form
factors were neither pinned down from data, nor were they estimated by the
resonance saturation principle. Instead, they were (on purpose) ignored
together with the analytic part of loop integrals, for details see \cite{B98}.
Even the incomplete result (\ref{axial}) used to be considered another example
of a rather slowly convergent $SU(3)$ chiral expansion.

Reordering of the perturbation series, which amounts here to the expansion
(\ref{LEC1}) of the LECs $d_{A}$ and $f_{A}$ , results again in a much
improved pattern
\begin{equation}%
\begin{array}
[c]{l}%
g_{1}\left[  pn\right]  \left(  0\right)  =1.30-0.14\\
g_{1}\left[  \Lambda\Sigma^{-}\right]  \left(  0\right)  =0.65-0.01\\
g_{1}\left[  p\Lambda\right]  \left(  0\right)  =-0.94-0.04\\
g_{1}\left[  \Lambda\Xi^{-}\right]  \left(  0\right)  =0.29+0.02\\
g_{1}\left[  n\Sigma^{-}\right]  \left(  0\right)  =0.30+0.02\\
g_{1}\left[  \Sigma^{0}\Xi^{-}\right]  \left(  0\right)  =0.92+0.13
\end{array}
\end{equation}
When compared with \cite{B98}, an impressive improvement is witnessed. Let us
recall that the reason of this improvement is simple. If any set of quantities
is relatively well described at two different chiral orders, then corrections
in the reordered perturbation series are small.

A dramatic change in the rate of convergence of the physical quantities is
reflected by a dramatic change in the values of LECs as pinned down at
different chiral orders. The values of $d_{A}$ and $f_{A}$ obtained at tree
level and the (incomplete) 1-loop level are different by a factor of two. It
was pointed out in \cite{JLMS92} that values of $d_{A}$ and $f_{A}$ as
obtained from the (incomplete) 1-loop fit reduce the large loop contributions
in the case of magnetic moments and are therefore preferable in loop
calculations. We emphasize that this philosophy is exactly opposite to the one
advocated here.

The problem with the approach \cite{JLMS92} is that the values of $d_{A}$ and
$f_{A}$ are probably to be changed again significantly when going one order
higher. In $SU(2)$ baryon ChPT, where the full 1-loop result for $g_{A} $ is
available \cite{KM99}, a large correction of about 25\% is present at the 4th
chiral order. This would change the whole numerical analysis of \cite{JLMS92}.
The advantage of the reordered ChPT is that it leaves lower orders untouched
after higher orders have been calculated and included.

\bigskip\ \ 

5. In conclusion, we have applied the reordering of ChPT, proposed by Becher
and Leutwyler \cite{BL99}, to the baryon octet static properties. We have
shown that the rate of convergence of the chiral expansion is significantly
(in some cases tremendously) improved. This does neither prevent large loop
corrections nor avoid large cancellations in higher orders. However, in the
reordered series the bulk of these large cancellations is due to the natural
cancellations caused by the reordering but not to the fine tuning of higher
order LECs.

The arguments given in this note are not sufficient to show that reordering of
the perturbation series always leads to a rapidly converging series. This
question depends on the process considered and has to be studied for further
applications in the future. However, we have shown that the examples
considered in \cite{DHB98} do not imply there is a problem with the rate of
convergence. Further processes which could shed light on the question
considered here include non-leptonic hyperon decays, hyperon polarizabilities,
strange form factors of the nucleon, Goldberger-Treiman discrepancy as well as
kaon photo-photo production. We plan to come back to these topics in a future publication.

\bigskip

\textbf{Acknowledgements. }We are grateful to Gerhard Ecker for careful
reading of the manuscript and useful comments.

\end{document}